\documentclass[authoryear]{FLO_v1}%

\usepackage{graphicx}
\usepackage{upgreek}
\usepackage{multicol,multirow}
\usepackage{amsmath,amssymb,amsfonts}
\usepackage{mathrsfs}
\usepackage{amsthm}
\usepackage[figuresright]{rotating}
\usepackage{appendix}
\usepackage[authoryear]{natbib}
\usepackage{ifpdf}
\usepackage[T1]{fontenc}
\usepackage{newtxtext}
\usepackage{newtxmath}
\usepackage{textcomp}
\usepackage{xcolor}

\usepackage[colorlinks,allcolors=blue]{hyperref}
\definecolor{jourcolor}{cmyk}{1,0.57,0.01,0.38}
\hypersetup{
    colorlinks,%
    citecolor=jourcolor,%
    filecolor=jourcolor,%
    linkcolor=jourcolor,%
    urlcolor=jourcolor
}

\theoremstyle{definition}

\articletype{RESEARCH ARTICLE}

\DOI{10.1017/flo.202X.X}

\Year{202X}

\Vol{X}

\Price{}

\art-id{FLO2000XXX}

\citearticle{Nishino, T. \& Smyth, A.S.M.}

\begin{document}

\title[Power loss mechanisms and optimal induction factors for realistic large wind farms]{Power loss mechanisms and optimal induction factors for realistic large wind farms}

\author[T. Nishino and A.S.M. Smyth]
{Takafumi Nishino$^{\ast}$ {{\href{https://orcid.org/0000-0001-6306-7702}{\includegraphics{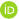}}}} and Amanda S. M. Smyth {{\href{https://orcid.org/0000-0002-4431-5915}{\includegraphics{orcid_logo}}}}
}

\address{Department of Engineering Science, University of Oxford, Oxford OX1 3PJ, UK}

\corres{*}{Corresponding author. E-mail:
\emaillink{takafumi.nishino@eng.ox.ac.uk}}


\date{\textbf{Received:} XX XX 2025; \textbf{Revised:} XX XX 202X; \textbf{Accepted:} XX XX 202X}

\abstract{
Power loss mechanisms in large wind farms are complex due to the multiscale nature of wind farm aerodynamics. Recent studies based on the two-scale momentum theory have brought new insights into this field; however, most of them have been limited to idealised wind farm scenarios. To better understand the power performance of real wind farms, in this study we extend the framework of the two-scale momentum theory to non-ideal turbine design and layout scenarios, and then introduce simple analytical sub-models to account for the associated power losses. These extensions provide a holistic view of how the turbine design, layout, operating conditions and atmospheric conditions collectively determine the amounts of different types of power losses in real wind farms, including the losses due to turbine-wake interference (i.e. `internal' power loss) and farm-atmosphere interaction (i.e. `external' power loss). We also present a simple iterative method for calculating the optimal farm induction factor that maximises the overall farm power for a given set of conditions, including the atmospheric boundary layer height. Analogously to the blade-element momentum theory playing a key role in wind turbine design optimisation, the present theory is expected to play a key role in wind farm design optimisation.
}

\maketitle

\begin{boxtext}

\textbf{\mathversion{bold}Impact Statement}

Whilst the installed capacity of wind power is increasing rapidly worldwide, there is no consensus on how the design and operation of large wind farms should be optimised. The difficulty lies in wind farm power losses being dependent on complex flow phenomena across a wide range of scales. The significance of this theoretical work is that it defines, and then predicts, all aerodynamic power losses in a wind farm using a closed set of physics-based algebraic equations, considering the first-order effects of turbine design, layout, operation and atmospheric conditions. This allows us to quickly estimate, among others, the relationship between the farm power and the farm induction factor, the latter of which has implications for wind farm wakes and their impacts on the surroundings. Therefore, this work provides a theoretical basis that can improve the process of wind farm design, deployment planning and environmental impact assessment in the future.
\end{boxtext}

\section{Introduction}

Wind farm aerodynamics is a relatively new and rapidly growing subject in applied fluid mechanics. In contrast to the fundamentals of rotor aerodynamics having been well explained (by the classical blade-element momentum theory of \citet{Gla1935}, for example) and sufficiently considered in the design of modern wind turbines, our understanding of wind farm aerodynamics is still immature, leaving some basic questions unresolved for today's wind power industry. The challenge in wind farm aerodynamics stems from its multiscale nature \citep{Vee19,PBS20} and the large number of parameters involved, ranging from regional weather conditions to the layout, design and operating conditions of individual turbines in the farm, resulting in a high-dimensional optimisation problem to consider. To provide a holistic view of wind farm performance, i.e. a physics-based prediction of how different types of power losses in a wind farm would change across the entire parameter space, it is necessary to develop a comprehensive theoretical model of wind farm aerodynamics.

A series of recent studies regarding the `two-scale momentum theory' \citep{N16,NH18,ND20,KND22,KDN23,KNL25} contributes to this goal. The first theoretical model proposed by \citet{N16} was similar to the classical `top-down' wind farm models \citep{Fra92,EF93,CMM10} with a key difference being that it introduced the concept of `farm-layer-average' wind speed to avoid considering the vertical profile of the atmospheric boundary layer (ABL) explicitly. The advantage of not explicitly considering the ABL profile was not obvious at the time, as these early models were for infinitely large wind farms and thus their applications to real wind farms were limited. Around the same time, \citet{SGM16} derived a coupled wake boundary layer (CWBL) model by coupling a traditional turbine-wake model \citep{Jen83,Kat86} with the top-down model of \citet{CMM10}, showing good agreement with large-eddy simulations (LES) of a finite-size wind farm; however, difficulties in comparing with real wind farm data remained, as some of the model input parameters were uncertain from field measurements. The advantage of not necessarily requiring the ABL profile as a model input became clearer when the concept of `momentum availability' was introduced by \citet{ND20}, generalising the two-scale theory to consider wind farms of different sizes, and subsequently its analytical sub-models by \citet{KDN23} predicting the momentum availability from limited information of the ABL (and the wind farm itself). As will be discussed later in this paper, the two-scale momentum theory provides a basic framework for the modelling of wind farm aerodynamics, upon which different types of sub-models (either analytical or numerical, with different levels of complexity depending on the type of information available for the model input) can be developed and employed to predict the performance of a given wind farm in a given environment.

Another key feature of the two-scale momentum theory is that it provides a different view of power loss mechanisms for large wind farms \citep{KND22,KNL25} from a traditional view based on the superposition of turbine-wake models \citep{Jen83,Kat86,BP14,Nyg20}. In short, the traditional view is that the (aerodynamic) power loss in a wind farm is due to individual turbine wakes reducing the inflow speed for downstream turbines. This traditional view has been slightly modified during the last decade as the effect of `wind farm blockage' \citep{AM17,WP17,Ble18} has been widely recognised and some analytical farm-blockage models \citep{Bra20,Nyg20,Seg21} have been developed. However, a common view in today's wind industry is still that the power loss in a single wind farm is mostly due to turbine wakes, as the reduction of inflow speed for front-row turbines due to farm blockage is usually (much) smaller than that for downstream turbines due to turbine wakes. In contrast to this, the view provided by the two-scale theory is that the power loss in a wind farm is primarily due to farm-scale flow interactions between the atmosphere and the wind farm as a whole, causing a reduction of `farm-average' wind speed (similarly to the concept of top-down models for infinitely large wind farms), while individual turbine-wake interactions within the farm may cause some additional power losses depending on the turbine layout. \citet{KNL25} have recently reported a detailed analysis of a large set of wind farm LES results to show that the `wind extractability' for a whole farm (i.e., {\it rate of change} of momentum availability for the whole farm {\it with respect to the farm-average wind speed}) is dependent on the farm length but insensitive to the turbine layout, suggesting that the view provided by the two-scale theory is more realistic unless the farm is too small. Note, however, that the new definitions of turbine-scale and farm-scale power losses introduced by \citet{KNL25} may be slightly misleading, as will be discussed later in Section \ref{sec:results}. 

The concept of farm-average wind reduction or `farm induction' may play a key role in predicting not only the power of a single farm but also the strength of `farm wake' and thus the power of another farm located downstream \citep{Lun19,SS22,Mey22}. The two-scale theory and its analytical sub-models allow very fast predictions of farm induction; however, the main interest of previous theoretical studies \citep{ND20,KDN23} was to predict an upper limit to the power of a single farm and thus the studies were mostly limited to `ideal' wind farm scenarios, considering aerodynamically ideal turbines (or actuator discs) in a hypothetical array where the inflow speed for each turbine was assumed to agree with the farm-average wind speed. In this study we extend the two-scale theory to account for the thrust and power characteristics of real turbine rotors, as well as the effects of arbitrary turbine layout and wind direction. We will also present a simple iterative method for calculating the optimal farm induction factor that maximises the power of a given wind farm. In addition to presenting these new results, another purpose of the present paper is to provide a concise and up-to-date description of the background theory for future reference.

%%%%%%%%%%%%%%%%%%%%%%%%%%%%%%%%%%%%%%%%%%%%

\section{Background theory}

\begin{figure}[t]
\centering{\includegraphics[width=34pc]{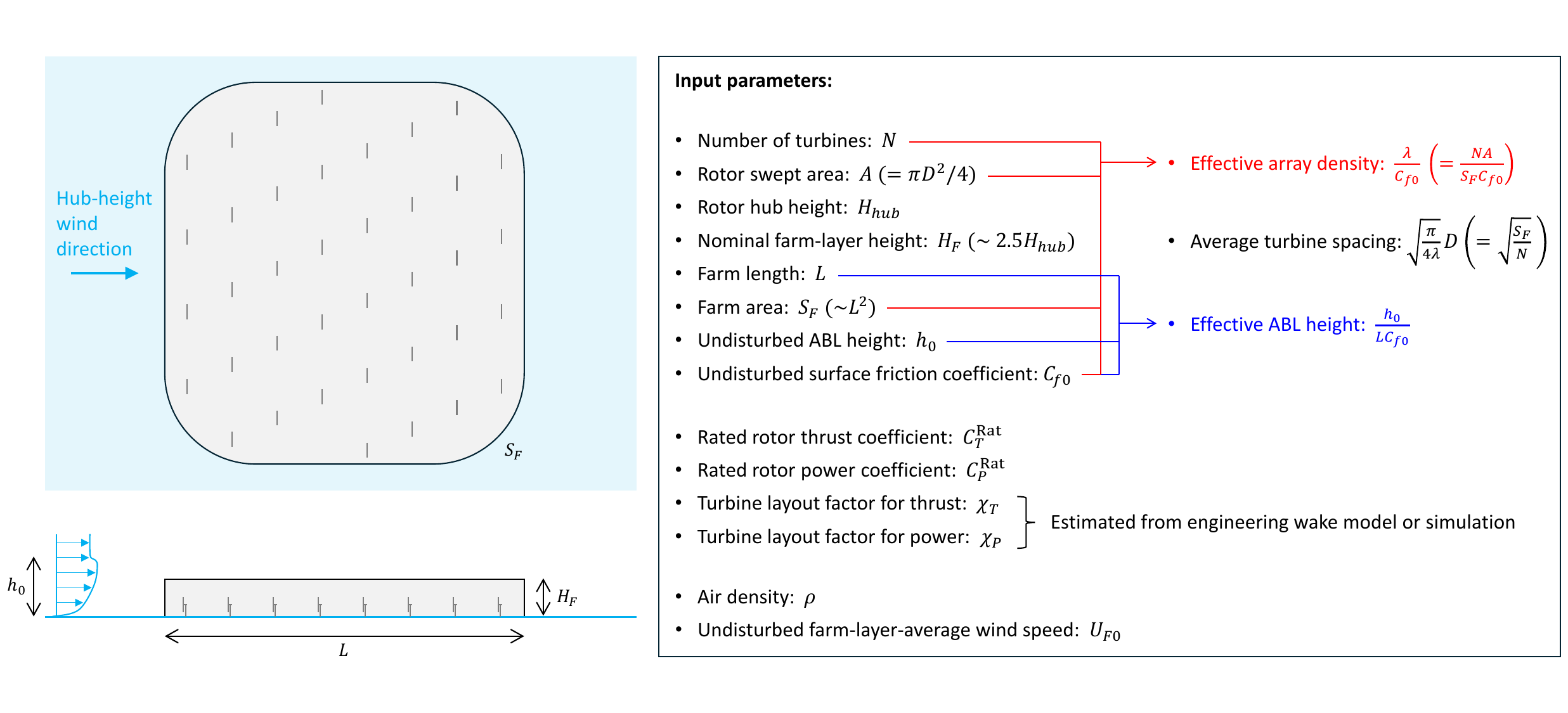}}\vspace{-9pt}
\caption{Schematic of flow configuration and a list of input parameters for the extended theoretical model. Note that, to calculate the non-dimensional farm power, only the non-dimensional parameters ($\lambda/C_{f0}, h_0/LC_{f0},C_T^{\rm Rat},C_P^{\rm Rat},\chi_T$ and $\chi_P$) are required as input. In addition to these parameters, one additional input (typically $C_T$ to obtain $\beta$, or vice versa) is required to solve the problem.}
\label{fig1}
\end{figure}

We summarise the existing two-scale momentum theory and  its analytical sub-models in this section, before describing the new additions in the next section. A schematic of the flow configuration and a list of input parameters for the present theoretical model (including the new additions) are shown in figure \ref{fig1}.

\subsection{Two-scale momentum theory} \label{sec:TSMT}
The core idea of the theory proposed by \citet{ND20} (hereafter referred to as ND20) is that a large wind farm is considered as a single power generation device (or an actuator volume) which imposes `farm-scale' resistance to the ABL flow and thus reduces `farm-average' wind speed as it generates power. Hence, the most important parameter describing the overall state of the farm (at a given time, i.e., averaged over a short period of time, typically 10 minutes) is its wind-speed reduction factor defined as
\begin{equation}
\beta\equiv \frac{U_F}{U_{F0}} ,
    \label{eqn:beta}
\end{equation}
where $U_F$ and $U_{F0}$ are the farm-average wind speeds for the cases with and without the farm present, respectively, and the farm induction factor can also be defined as $b\equiv (1-\beta)\equiv (U_{F0}-U_F)/U_{F0}$. This is analogous to the classical actuator disc theory for a single turbine, where the state of airflow through the turbine is described using the axial induction factor defined as $a\equiv (U_{T0}-U_T)/U_{T0}$, where $U_T$ and $U_{T0}$ are `turbine-average' wind speeds for the cases with and without the turbine present, respectively. In reality, neither the flow through a turbine nor the flow through a farm is perfectly uniform. Nonetheless, the average wind speed reduction is the key factor in each problem as it captures the first-order effect of power generation on the undisturbed wind, which in turn determines the amount of wind power extractable by a device (turbine or farm) and the significance of its wake.

A key question in the two-scale momentum theory is how to define the nominal `farm-layer height' across which the farm-average wind speeds $U_F$ and $U_{F0}$ are calculated. Unlike the actuator disc theory for a single turbine, where the rotor swept area $A$ is an obvious choice for the area across which the turbine-average wind speeds $U_T$ and $U_{T0}$ are calculated, there is no obvious choice for the farm-layer height in the two-scale theory. Essentially, different choices of the farm-layer height lead to different values of `reference' wind speed $U_{F0}$ for non-dimensionalisation. The original definition of the farm-layer height $H_F$ adopted by ND20 is such that it gives $U_{F0}=U_{T0}$. This definition leads to a concise description of power loss mechanisms as shown later, although the vertical profile of the undisturbed wind speed, $U_0(z)$, is required to calculate $H_F$. When the vertical profile of the ABL is unknown, the exact value of $H_F$ (that gives $U_{F0}=U_{T0}$) cannot be calculated; however, the concept of farm-layer average (across an unknown height $H_F$) remains useful. As described later in Section 2.2, \citet{KDN23} have derived an analytical sub-model of the two-scale theory to predict the farm-layer-average wind reduction factor $\beta$ without using $H_F$ as input. It is also worth noting that $H_F \approx 2.5H_{hub}$ (where $H_{hub}$ is the hub height of turbine rotors) is a good approximation for a range of turbine designs and ABL profiles \citep{KND22}.

The essence of the two-scale momentum theory lies in the expression of the non-dimensional farm momentum (NDFM) equation, derived from the law of conservation of linear momentum for the cases with and without the farm present (ND20, see also \citet{KND22}). The final form of the NDFM equation, from which the wind reduction factor $\beta$ can be calculated, is
\begin{equation}
    C_T^* \frac{\lambda}{C_{f0}} \beta^2 + \beta^\gamma = M,
    \label{eqn:NDFM}
\end{equation}
where the first and second terms on the left-hand side represent the total turbine drag ($\Sigma_{i=1}^N T_i$, where $N$ is the number of turbines in the farm) and surface friction drag ($\tau_w S_F$, where $S_F$ is the horizontal area of the farm) normalised by the `undisturbed' surface friction drag ($\tau_{w0}S_F$), respectively. Note that $C_T^*\equiv \Sigma_{i=1}^N T_i/\frac{1}{2}\rho U_F^2NA$ is the `internal' thrust coefficient (i.e. defined using $U_F$ instead of $U_{F0}$), $\lambda\equiv NA/S_F$ is the array density, $C_{f0}\equiv \tau_{w0}/\frac{1}{2}\rho U_{F0}^2$ is a surface friction coefficient, and $\gamma$ is a surface friction exponent. Another term could be added to the left-hand side to account for the drag caused by turbine support structures separately \citep{MNA19}; however, this drag is usually much smaller than the drag due to turbine rotors. The right-hand side, called the momentum availability factor, $M$, represents the amount (per unit time) of momentum in the hub-height wind direction supplied by the atmosphere to the wind farm site ($X_F$) normalised by that for the undisturbed case ($X_{F0}$), i.e.,
\begin{equation}
    M\equiv \frac{X_F}{X_{F0}} = \frac{X_{\textrm{adv}}+X_{\textrm{dif}}+X_{\textrm{pgf}}+X_{\textrm{Cor}}+X_{\textrm{uns}}}{X_{\textrm{adv,0}}+X_{\textrm{dif,0}}+X_{\textrm{pgf,0}}+X_{\textrm{Cor,0}}+X_{\textrm{uns,0}}},
    \label{eqn:Mdefinition}
\end{equation}
where $X_{\textrm{adv}}$, $X_{\textrm{dif}}$, $X_{\textrm{pgf}}$, $X_{\textrm{Cor}}$ and $X_{\textrm{uns}}$ are the {\it net rates} of momentum transfer due to advection, diffusion (stress), pressure gradient force, Coriolis force, and unsteadiness (local acceleration/deceleration), respectively, and the subscript 0 denotes the undisturbed case. Note that $X_F=\Sigma_{i=1}^N T_i + \tau_w S_F$ and $X_{F0}=\tau_{w0}S_F$ due to the momentum balance for the cases with and without the farm, respectively.

The NDFM equation (\ref{eqn:NDFM}) expresses the farm-scale momentum balance in a concise manner. When there is no turbine in a given farm area ($\lambda=0$), the equation becomes `$0+1=1$', meaning that the momentum supplied by the atmosphere per unit time (right-hand side) is balanced by the surface friction (second term on the left-hand side). For a typical offshore wind farm, the second term often decreases from 1 down to about 0.6 to 0.9 as the farm-average wind speed decreases ($\beta<1$), whilst the first term often increases up to about 5 to 10 \citep{KDN23,KNL25}, meaning that the turbine drag is about 5 to 10 times larger than the undisturbed surface friction drag (depending largely on the array density $\lambda$ and the momentum availability factor $M$, i.e., how the atmosphere responds to the wind farm). Since the turbine drag is usually much larger than the surface friction drag, the wind reduction factor $\beta$ is usually not sensitive to the value of $\gamma$ for offshore wind farms, allowing us to assume $\gamma=2.0$ for convenience, although this may slightly overestimate $\beta$ and the farm power (ND20).

The usefulness of the two-scale momentum theory becomes clearer when the atmospheric response to the wind farm is quantified by introducing a parameter $\zeta$, which was originally called the momentum response factor (ND20) but then renamed the wind extractability factor \citep{KND22} (to avoid confusion with the momentum availability factor). This parameter $\zeta$ represents the {\it rate of change} of the momentum availability factor {\it with respect to the farm induction factor}. Specifically, $\zeta$ is defined by
\begin{equation}
    M = 1 + \zeta(1-\beta),
    \label{eqn:Mtozeta}
\end{equation}
where, in general, $\zeta$ itself is also a function of $\beta$, i.e. the relationship between $M$ and $\beta$ is nonlinear. However, for large offshore wind farms where $\beta$ is in a typical range of about 0.75 to 1, the relationship between $M$ and $\beta$ is expected to be approximately linear (ND20, see also \citet{PDN21}). Recent LES results by \citet{KNL25} show that $\zeta$ is sensitive to the ABL characteristics but insensitive to the turbine layout and the value of $\beta$, further supporting the linear approximation of $M$ against $\beta$ for large offshore wind farms. This approximate linearlity suggests that the two-scale momentum theory, featuring the NDFM equation with the concepts of farm induction, momentum availability  and wind extactability factors, may allow us to significantly reduce the {\it dimension} of the problem of wind farm aerodynamics by splitting the problem into `internal' (turbine-scale) and `external' (farm-scale) sub-problems. The former is mainly to predict $C_T^*$ on the left-hand side and the latter is to predict $\zeta$ on the right-hand side of the NDFM equation, from which the farm-average wind speed (and eventually the farm power) can be predicted in a `loosely' coupled manner (ND20).

Note that the core framework of the two-scale theory, consisting of the above equations (\ref{eqn:beta}) to (\ref{eqn:Mtozeta}), is generic in the sense that it has been derived from the conservation of linear momentum for general three-dimensional (3D) flow over a wind farm (equivalent to the streamwise momentum equation of the unsteady 3D Reynolds-averaged Navier-Stokes equations) with little assumptions on the farm design or the atmospheric flow conditions. The only assumptions adopted are: (i) the farm is on a flat terrain or sea surface, and (ii) the farm-average air density $\rho$ is not affected by the farm (i.e. the change of $\rho$ due to possible farm-induced changes in air temperature and humidity is negligibly small). Hence, the theory may help a wide range of studies on wind farm aerodynamics, not only for analytical modelling of wind farm power production (as described in the rest of the paper) but also for the development of multiscale-coupled computational models, parameterisation of large wind farms in numerical weather models, and the analysis of wind farm flow in general.

\subsection{Modelling of $M$}
To evaluate the momentum availability factor $M$, we need to consider a large control volume (CV) that contains the entire wind farm, like the grey domain shown in figure \ref{fig1}. Note that the height of the CV, for which the momentum balance is considered, must be large enough to contain all turbines (i.e. it must be taller than $H_{hub}+D/2$, where $H_{hub}$ is the rotor hub height and $D$ the rotor diameter) but this does not need to be the same as the farm-layer height $H_F$. Depending on the choice of the CV height, the balance between different terms in $M$, namely $X_{\textrm{adv}}$, $X_{\textrm{dif}}$, $X_{\textrm{pgf}}$, $X_{\textrm{Cor}}$ and $X_{\textrm{uns}}$ in equation (\ref{eqn:Mdefinition}), may change, but the value of $M$ itself is not affected by this choice (as long as it is taller than $H_{hub}+D/2$). However, as shown by \citet{KDN23} (hereafter referred to as KDN23), it is usually convenient to set the CV height to be $H_F$ (even if the exact value of $H_F$ is unknown) since we usually need to model $M$ as a function of $\beta$ so that the NDFM equation (\ref{eqn:NDFM}) can be solved for $\beta$.

The analytical model of $M$ proposed by KDN23 is for a CV of height $H_F$, which is assumed to be smaller than the undisturbed ABL height $h_0$ as illustrated in figure \ref{fig1} (note that this assumption may not always hold for very tall wind turbines in recent years). This model is for quasi-steady scenarios, i.e. $X_{\textrm{uns}}=X_{\textrm{uns,0}}=0$. It is also assumed that the contribution of the Coriolis force to the momentum balance in the hub-height wind direction is negligibly small, i.e. $X_{\textrm{Cor}}=X_{\textrm{Cor,0}}=0$. Hence, the equation (\ref{eqn:Mdefinition}) is simplified to
\begin{equation}
    M = 1 + \Delta M_{\textrm{adv}} + \Delta M_{\textrm{dif}} + \Delta M_{\textrm{pgf}},
    \label{eqn:Mdecompose}
\end{equation}
where $\Delta M_{\textrm{adv}}=(X_{\textrm{adv}}-X_{\textrm{adv,0}})/X_{F0}$, $\Delta M_{\textrm{dif}}=(X_{\textrm{dif}}-X_{\textrm{dif,0}})/X_{F0}$, and $\Delta M_{\textrm{pgf}}=(X_{\textrm{pgf}}-X_{\textrm{pgf,0}})/X_{F0}$. Using quasi-one-dimensional (1D) flow assumptions, KDN23 derived a simple algebraic model for the sum of the advection and pressure terms as
\begin{equation}
    \Delta M_{\textrm{adv}} + \Delta M_{\textrm{pgf}} = \frac{H_F}{LC_{f0}}\left(1-\beta^2 \right),
    \label{eqn:Madvpgf}
\end{equation}
where $L$ is the streamwise length of the farm. KDN23 also derived a simple model for the diffusion term, assuming fully-developed and self-similar shear stress profiles to evaluate the rate of momentum entrainment across the top surface of the CV (and ignoring any momentum diffusion across the sides of the CV), as
\begin{equation}
    \Delta M_{\textrm{dif}} = M\left(1-\frac{H_F}{h_0}\beta \right) - \left(1-\frac{H_F}{h_0} \right),
    \label{eqn:Mdif}
\end{equation}
where the following two approximations had been adopted (see KDN23 for details):
\begin{equation}
    1-\frac{\tau_{t0}}{\tau_{w0}} \approx \frac{H_F}{h_0},
    \label{eqn:Hf/h0}
\end{equation}
where $\tau_{t0}$ is the undisturbed shear stress at the top of the CV ($z=H_F$), and
\begin{equation}
    \frac{h}{h_0} \approx \frac{1}{\beta},
    \label{eqn:h/h0}
\end{equation}
where $h$ is the ABL height for the case with the farm present. It should be noted that the model of $\Delta M_{\textrm{dif}}$, which is the diffusion term of $M$, contains $M$ itself (equation (\ref{eqn:Mdif})). This means that $\Delta M_{\textrm{dif}}$ is actually modelled in a coupled manner with $\Delta M_{\textrm{adv}}$ and $\Delta M_{\textrm{pgf}}$ in equation (\ref{eqn:Madvpgf}), even though they appear as separate terms in equaition (\ref{eqn:Mdecompose}). Combining equations (\ref{eqn:Mdecompose}) to (\ref{eqn:Mdif}), we obtain
\begin{equation}
    M = \frac{1+\frac{h_0}{LC_{f0}}(1-\beta^2)}{\beta},
    \label{eqn:Mmodel}
\end{equation}
which is arguably the most basic model of $M$ to capture the first-order effects of the undisturbed ABL height $h_0$ and the farm length $L$. Note that both $\Delta M_{\textrm{adv}} + \Delta M_{\textrm{pgf}}$ (equation (\ref{eqn:Madvpgf})) and $\Delta M_{\textrm{dif}}$ (equation (\ref{eqn:Mdif})) depend on the CV height $H_F$, but the resulting model of $M$ (equation (\ref{eqn:Mmodel})) is not dependent on $H_F$ as the effects of $H_F$ cancel each other out. Hence, this model satisfies the requirement that the choice of the CV height does not affect the value of $M$. This model also captures the important trend that the momentum supplied by the atmosphere to the farm becomes more and more due to diffusion (vertical turbulent entrainment) as the farm size increases, since $\Delta M_{\textrm{adv}}+\Delta M_{\textrm{pgf}}$ decreases as $L$ increases.

The simplicity of the above model proposed by KDN23 is justified since, in real-world applications, the exact profile of the undisturbed ABL profile is usually unknown and unavailable as input. However, when good estimates of $\tau_{t0}/\tau_{w0}$ and $H_F$ are available as input, e.g. from numerical weather prediction (NWP) simulations and/or measurements, the above model of $M$ can be improved by avoiding the use of equation (\ref{eqn:Hf/h0}), which often causes a large error in $M$ as it approximates $\tau_{t0}/\tau_{w0}$ by assuming a linear increase in shear stress from the top ($z=h_0$) to the bottom ($z=0$) of the undisturbed ABL. It should also be noted that, in reality, it is often a challenge to even get a good estimate of $h_0$. This suggests that, in future studies, an empirical correction factor could be proposed and applied to the value of $h_0$ in the above equations (\ref{eqn:Mdif}) to (\ref{eqn:Mmodel}) to account for uncertainty or bias in the estimation of $h_0$ as well as for any errors arising from the self-similar approximation of the shear stress profile.

Finally, when the wind reduction factor is in the range of $0.8\leq\beta\leq 1$, the model of $M$ in equation (\ref{eqn:Mmodel}) can be simplified even further, using a linear approximation (see KDN23 for details), to obtain
\begin{equation}
    M \approx 1+\left(1.18+2.18\frac{h_0}{LC_{f0}} \right)(1-\beta), \hspace{12pt}
    \zeta \approx 1.18+2.18\frac{h_0}{LC_{f0}}.
    \label{eqn:Mapprox}
\end{equation}

\subsection{Modelling of $C_T^*$}
To solve the NDFM equation (\ref{eqn:NDFM}) for $\beta$ we also need to model the `internal' thrust coefficient $C_T^*$, for which a few different methods have been proposed in the past. The first model of $C_T^*$ was proposed by \citet{N16} (see also \citet{KND22}) for an infinitely large array of ideal turbines, expressed as $C_T^*=16C_T'/(4+C_T')^2$, where $C_T'$ is the resistance coefficient of the turbines. This model was derived from the classical actuator disc theory with the assumption that the `inflow' speed (or `upstream' wind speed) for each turbine in a farm is the same as the farm-average wind speed. Recent LES results for a finite array of actuator discs show that this simple model works well for ideal turbines arranged in a staggered manner, but over-predicts $C_T^*$ (and thus the farm power) when the turbines are aligned with the wind direction \citep{KDN23,KNL25}. Another model of $C_T^*$ has been proposed by \citet{NH18}, who adopted the blade-element momentum (BEM) theory to account for the effects of the design and operating conditions of real turbine rotors, such as the blade pitch angle and the tip-speed ratio, but the same assumption for the inflow speed was used to neglect the turbine layout effect. More recently, \citet{LPN23} and \citet{PNK24} used an engineering turbine-wake model to predict the turbine layout effect on $C_T^*$ numerically instead of analytically.

%%%%%%%%%%%%%%%%%%%%%%%%%%%%%%%%%%%%%%%%%%%%

\section{New additions to the theory}
\label{sec:newTheory}

\subsection{Power loss mechanisms} \label{sec:powerloss}
We consider that the power losses in a large wind farm can be classified into the following three types: (i) loss due to turbine design (relative to the power of ideal turbines, i.e. actuator discs); (ii) loss due to `turbine-scale' or `internal' flow interactions (i.e. direct interference of turbine wakes with downstream turbines, reducing the inflow speed for those turbines relative to the farm-average wind speed); and (iii) loss due to `farm-scale' or `external' flow interactions (i.e. interaction of the whole wind farm with the atmosphere, reducing the farm-average wind speed). To define these power losses mathematically, we introduce three different types of farm power coefficients:

\begin{equation}
    C_{PG} \equiv \frac{\langle P \rangle}{\frac{1}{2}\rho U_{F0}^3 A},\hspace{12pt}
    C_{P}^* \equiv \frac{\langle P \rangle}{\frac{1}{2}\rho U_{F}^3 A},\hspace{12pt}
    C_{P} \equiv \frac{\langle P \rangle}{\frac{1}{2}\rho \langle U_{T,in}^3\rangle A},
    \label{eqn:3CP} 
\end{equation}
where $\langle P \rangle \equiv \Sigma_{i=1}^N P_i/N$ is the farm-averaged value of the turbine power. The first power coefficient, $C_{PG}$, is the `global' power coefficient of the farm, using $U_{F0}^3$ to non-dimensionalise the power. Note that $U_{F0}$ is the `undisturbed' wind speed, which is typically used in resource assessments. The second one, $C_P^*$, is the `internal' power coefficient of the farm, using $U_{F}^3$ instead of $U_{F0}^3$. The third one, $C_P$, is the `effective' power coefficient of all individual turbines in the farm, using $\langle U_{T,in}^3 \rangle \equiv \Sigma_{i=1}^N (U_{T,in}^3)_i/N$ for non-dimensionalisation, where $U_{T,in}$ is the inflow speed for a given turbine. In reality, this inflow speed $U_{T,in}$ is difficult to find in a wind farm, as it is difficult to know where the `upstream' position of each turbine really is. Nonetheless, this $U_{T,in}$ is equivalent to the upstream wind speed for an isolated turbine, i.e. this is the (unknown) upstream wind speed with which the thrust and power coefficients of a given turbine in a wind farm will agree with those of the same turbine in isolation. Hence, in special cases where all $N$ turbines in a farm have the same power and the same inflow speed, this $C_P$ becomes identical to the traditional power coefficient of a single turbine.

Now we define the three types of wind farm power losses using the three power coefficients defined above. First, the difference between $C_{PG}$ and $C_P^*$ represents the loss due to `external' flow interactions between the whole wind farm and the atmosphere. Since the actual power of a given wind farm is $N\langle P \rangle =\frac{1}{2}\rho U_{F0}^3 ANC_{PG}$ whereas the power of a hypothetical farm {\it without power loss due to external flow interactions} would be $\frac{1}{2}\rho U_{F0}^3 ANC_{P}^*$, we define the `external efficiency' of a wind farm as
\begin{equation}
    \eta_{\rm ext} \equiv \frac{\frac{1}{2}\rho U_{F0}^3 ANC_{PG}}{\frac{1}{2}\rho U_{F0}^3 ANC_{P}^*} = \frac{C_{PG}}{C_P^*} = \frac{U_F^3}{U_{F0}^3} = \beta^3 .
    \label{eqn:etaext} 
\end{equation}
Similarly, the difference between $C_P^*$ and $C_P$ represents the power loss due to `internal' flow interactions. Since the power of a hypothetical farm {\it without power loss due to internal or external flow interactions} would be $\frac{1}{2}\rho U_{F0}^3 ANC_{P}$, we define the `internal efficiency' of a wind farm as
\begin{equation}
    \eta_{\rm int} \equiv \frac{\frac{1}{2}\rho U_{F0}^3 ANC_{P}^*}{\frac{1}{2}\rho U_{F0}^3 ANC_{P}} = \frac{C_P^*}{C_P} = \frac{\langle U_{T,in}^3\rangle}{U_{F}^3} .
    \label{eqn:etaint} 
\end{equation}
Finally, the power loss due to non-ideal turbine design can be calculated from the difference between $C_P$ and the power coefficient of ideal turbines obtained from the classical actuator disc theory, $C_{P,{\rm ADT}}$ (see Section \ref{sec:phiP} for further details). Since the total power of $N$ ideal turbines in isolation would be $\frac{1}{2}\rho U_{F0}^3 ANC_{P,{\rm ADT}}$, we define the `rotor efficiency' as
\begin{equation}
    \eta_{\rm rot} \equiv \frac{\frac{1}{2}\rho U_{F0}^3 ANC_P}{\frac{1}{2}\rho U_{F0}^3 ANC_{P,{\rm ADT}}} = \frac{C_P}{C_{P,{\rm ADT}}}.
    \label{eqn:etarot} 
\end{equation}
Hence, the overall farm efficiency (i.e. the ratio of the actual farm power to the total power of the same number of ideal turbines in isolation) is $\eta_{\rm farm} = \eta_{\rm ext}\eta_{\rm int}\eta_{\rm rot}$.

The definition of $\eta_{\rm ext}$ in (\ref{eqn:etaext}) is very similar to the `farm-scale efficiency' introduced by \citet{KNL25} for wind farms with ideal turbines. However, \citet{KNL25} proposed to calculate this from the value of $\beta$ for a special `near-ideal' case (as will be discussed further in Section \ref{sec:results}). In contrast, our aim here is to calculate $\eta_{\rm ext}$ in (\ref{eqn:etaext}) for a {\it real} wind farm of interest by solving the NDFM equation (\ref{eqn:NDFM}) for $\beta$ for that farm (rather than for an idealised farm). To do this, we need a model to account for the effect of the turbine layout on $C_T^*$ in (\ref{eqn:NDFM}). In addition, to calculate $\eta_{\rm int}$ in (\ref{eqn:etaint}), we need a model of $C_P^*$ that accounts for the turbine layout effect. Hence, we introduce `turbine layout factors' as
\begin{equation}
    \chi \equiv \frac{\langle U_{T,in}\rangle}{U_F},\hspace{12pt}
    \chi_T \equiv \frac{\langle U_{T,in}^2\rangle}{U_F^2} = \frac{C_T^*}{C_T},\hspace{12pt}
    \chi_P \equiv \frac{\langle U_{T,in}^3\rangle}{U_F^3} = \frac{C_P^*}{C_P},
    \label{eqn:chi} 
\end{equation}
where $\chi_T$ and $\chi_P$ are the layout factors to be modelled for the thrust and power, respectively, and $C_T$ is the `effective' thrust coefficient of all individual turbines in the farm, i.e. $C_T \equiv \langle T \rangle / \frac{1}{2}\rho \langle U_{T,in}^2\rangle A$. Note that $\chi_T$ affects $\beta$ (via the NDFM equation) and thus $\eta_{\rm ext}$, whereas $\chi_P$ decides $\eta_{\rm int}$, and in general, the relationships between $\chi$, $\chi_T$ and $\chi_P$ are unknown. However, in special cases where all turbines have approximately the same inflow speed, we obtain $\chi_T\approx\chi^2$ and $\chi_P\approx\chi^3$, and hence $\chi_P\approx\chi_T^{3/2}$.

The definitions of the power coefficients, efficiencies and layout factors introduced above are generic and compatible with the core framework of the two-scale momentum theory summarised in Section \ref{sec:TSMT}, requiring no assumption on the spatio-temporal variation of the flow field (such as scale separation).

\subsection{Modelling of turbine layout factors} \label{sec:chimodel}
As noted earlier, recent studies \citep{LPN23,PNK24} show that an engineering wake model can be used to predict turbine layout effects on $C_T^*$ numerically for a given wind farm. However, since our aim here is to provide a holistic view of the power performance of {\it any} wind farms, we need an analytical model that allows us to predict $C_T^*$ (and $C_P^*$) instantly. In the following, we propose a basic method to derive such a model from an existing engineering wake model and LES data. Note that the two-scale separation assumption (ND20, see also \citet{KNL25}) is adopted here and in Section~\ref{sec:phiP}, i.e. we assume that $\chi$, $\chi_T$, $\chi_P(=\eta_{\rm int})$ and $\eta_{\rm rot}$ are `internal' parameters and therefore not directly affected by `external' flow conditions, such as the ABL height.

\begin{figure}[t]
\centering{\includegraphics[width=34pc]{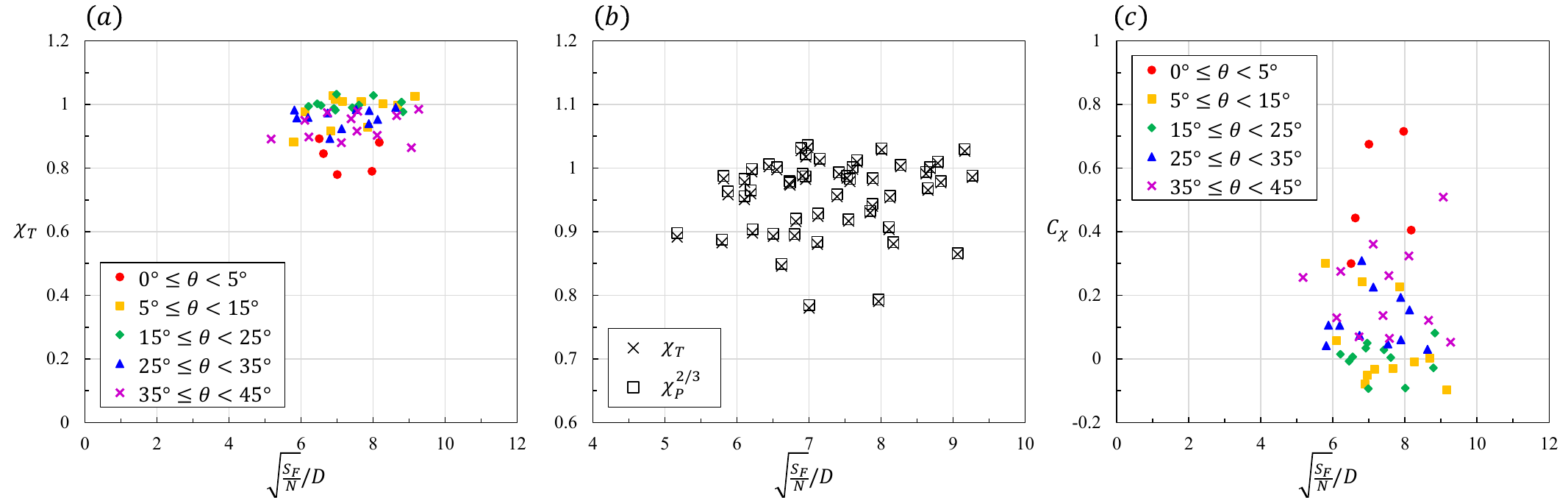}}\vspace{-2pt}
\caption{Turbine layout factors and the model coefficient $C_{\chi}$ computed from the LES results reported by \citet{KND22} for 50 different periodic arrays of actuator discs: (a) $\chi_T$ plotted against the non-dimensional average turbine spacing for five different sub-ranges of wind direction, (b) comparison of $\chi_T$ and $\chi_P^{2/3}$, and (c) the coefficient $C_{\chi}$ for the proposed model, equation (\ref{eqn:chimodel}).}
\label{fig2}
\end{figure}

\citet{KND22} performed 50 different cases of LES of a neutral ABL over a regular periodic array of actuator discs with $C_T'=1.33$ (which corresponds to $C_T=0.75$) for a range of turbine spacing ($5D\le s_x \le10D$ and $5D \le s_y \le 10D$) and wind direction ($0^{\circ} \le \theta < 45^{\circ}$). The values of $\chi_T$ computed from the 50 LES cases are plotted in figure \ref{fig2}($a$) against the non-dimensional average turbine spacing $\sqrt{s_xs_y}/D=\sqrt{S_F/ND^2}$ for five sub-ranges of $\theta$. It is clear that $\chi_T$ is close to 1 for the majority of the 50 cases, but decreases to about 0.8 when $\theta$ is close to $0^{\circ}$, i.e. when the wind direction is close to one of the two axes of the regular array, causing significant wake interference. These $\chi_T$ values are almost identical to the values of $\chi_P^{2/3}$ as shown in figure \ref{fig2}($b$) for the same 50 LES cases, since the inflow speed $ U_{T,in}$ is almost the same for all turbines in each regular periodic array simulated. Although real wind farms may have an irregular array of turbines, here we assume $\chi_T = \chi^2$ and $\chi_P = \chi^3$ for simplicity and propose a single model of $\chi$ that can reproduce the layout effects observed in these LES data.

The LES results plotted in figure \ref{fig2}($a$) do not show a clear dependency of $\chi_T$ on the turbine spacing, but theoretically $\chi$, $\chi_T$ and $\chi_P$ should all converge to 1 as the turbine spacing increases and all flow interaction effects diminish. Also, these LES results are for a given $C_T$ value (0.75) but $\chi$, $\chi_T$ and $\chi_P$ should all converge to 1 as $C_T$ approaches zero. Since the dependency of $\chi$ on $C_T$ and turbine spacing is expected to be similar to how the wind speed behind a single turbine depends on its thrust and streamwise distance, we adopt the formulation of the traditional wake model of \citet{Jen83} and \citet{Kat86} and introduce a model coefficient $C_\chi$ to formulate our analytical model of $\chi$ as
\begin{equation}
    \chi = \chi_T^{1/2} = \chi_P^{1/3} = 1-C_{\chi}\left[ \frac{1-\sqrt{1-C_T}}{( 1+2k\sqrt{\pi/4\lambda} )^2} \right],
    \label{eqn:chimodel} 
\end{equation}
where the term inside the square brackets represents the `wake deficit' expected at a distance $D\sqrt{\pi/4\lambda}$ downstream of a given turbine, and $k=0.05$ is a typical wake growth rate \citep{PBS20}. Note that $D\sqrt{\pi/4\lambda}=\sqrt{S_F/N}=\sqrt{s_xs_y}$ is the average turbine spacing in a farm. This means that $C_\chi \approx 1$ is expected when the wind direction is aligned with an axis of a regular array of turbines with $s_x=s_y$, whereas $C_\chi \approx 0$ is expected when the wind direction is such that there is little wake interference. This is confirmed by calculating $C_\chi$ directly from the LES results, i.e. by substituting the LES results of $\chi_T$ shown in figure \ref{fig2}($a$) into equation (\ref{eqn:chimodel}) together with the values of $C_T$ and the average turbine spacing. The $C_\chi$ values calculated are plotted in figure \ref{fig2}($c$) for all 50 LES cases, showing that $C_\chi<0.5$ for most cases (since the wind direction is not aligned for most cases). Note that $C_{\chi}$ may take a small negative value since, depending on the wind direction, the average turbine inflow speed $\langle U_{T,in}\rangle$ may become slightly higher than $U_F$ due to local flow acceleration; see \citet{KND22} for further details.

Although $C_{\chi}$ depends significantly on the wind direction, in this study we will adopt (unless stated otherwise) $C_{\chi}=0.14$, the average value of all 50 LES results shown in figure \ref{fig2}($c$). Note that these 50 cases cover a typical range of layout parameter space ($s_x$, $s_y$, $\theta$) in an unbiased manner, representing a general set of layout scenarios; hence, this $C_{\chi}=0.14$ is our estimate of the (long-time) average value of $C_{\chi}$ for typical offshore wind farms. In reality, however, it is common that there is a dominant wind direction at a given site and wind farm developers ensure that the axes of their turbine array are not aligned with that direction, meaning that the average value of $C_{\chi}$ could be lower. In addition, some flow control techniques to reduce turbine-wake interference, such as `wake steering' \citep{Fle17,HLD19}, could lower the average value of $C_{\chi}$ even further.

\subsection{Modelling of rotor efficiency} \label{sec:phiP}
\begin{figure}[t]
\centering{\includegraphics[width=34pc]{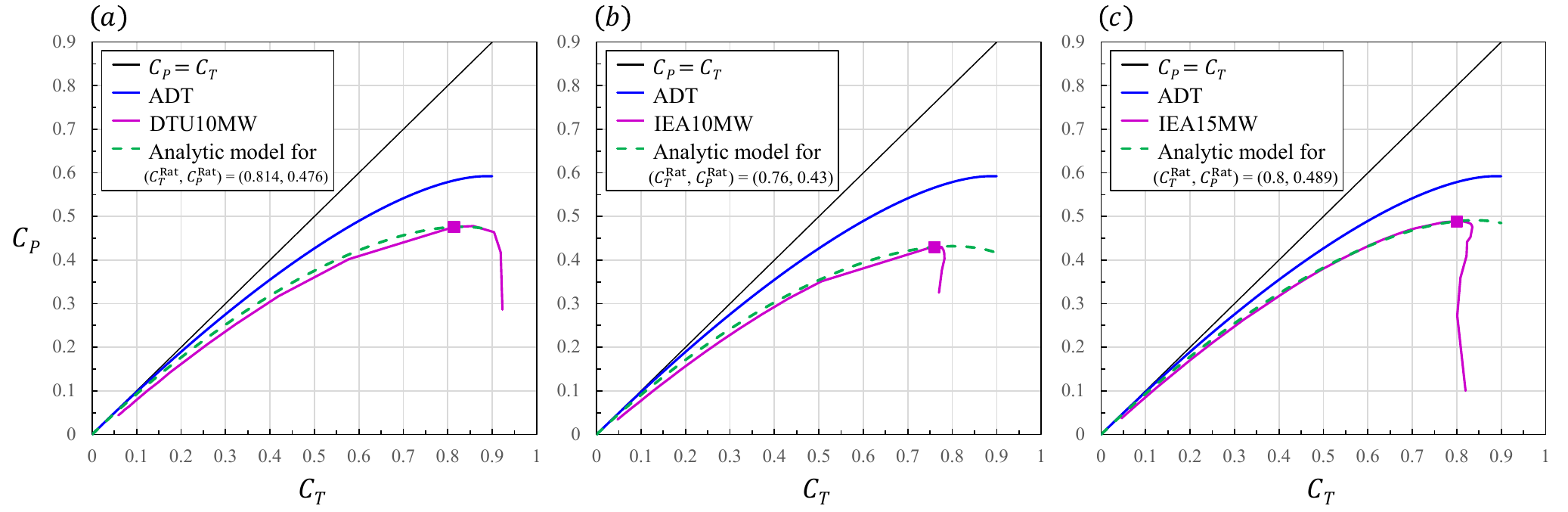}}\vspace{-0pt}
\caption{Comparison of standard $C_P$ vs $C_T$ curves for three reference wind turbine models (purple) with the classical ADT (blue) and the proposed analytical model (green): (a) DTU 10 MW turbine, (b) IEA 10 MW turbine, and (c) IEA 15 MW turbine. The square symbols show the rated operating points.}
\label{fig3}
\end{figure}

The rotor efficiency $\eta_{\rm rot}$ can be predicted by the classical BEM theory as a function of the rotor design and operating conditions (such as the blade pitch angle and the tip-speed ratio). Using the BEM theory together with the two-scale momentum theory would enable optimisation of the design and operation of wind turbines and wind farms in a coupled manner \citep{NH18}. However, since our focus here is on wind farm performance and we prefer to keep this theoretical farm model as simple as possible, we introduce an analytical model of $\eta_{\rm rot}$ that requires only the rated thrust coefficient, $C_T^{\rm Rat}$, and rated power coefficient, $C_P^{\rm Rat}$, of real turbines as input. Specifically, we use $C_T^{\rm Rat}$ and $C_P^{\rm Rat}$ to correct the power coefficient of aerodynamically ideal turbines, $C_{P,{\rm ADT}}$, to estimate $C_P$ as follows:
\begin{equation}
    C_P = \eta_{\rm rot}C_{P,{\rm ADT}} = \left[ \frac{\sigma C_P^{\rm Rat}+(1-\sigma) C_{P,{\rm ADT}}^{\rm Rat}}{C_{P,{\rm ADT}}^{\rm Rat}}\right] C_{P,{\rm ADT}},
    \label{eqn:CPmodel} 
\end{equation}
where
\begin{equation}
    C_{P,{\rm ADT}} = \frac{1}{2}C_T\left(1+\sqrt{1-C_T} \right), \hspace{10pt}
    C_{P,{\rm ADT}}^{\rm Rat} = \frac{1}{2}C_T^{\rm Rat}\left(1+\sqrt{1-C_T^{\rm Rat}} \right), \hspace{10pt}
    \sigma = \left( \frac{\frac{C_T}{C_{P,{\rm ADT}}}-1}{\frac{C_T^{\rm Rat}}{C_{P,{\rm ADT}}^{\rm Rat}}-1} \right)^{\frac{1}{2}}.
    \label{eqn:CPADT} 
\end{equation}
Here the model parameter $\sigma$ has been designed to increase exponentially from 0 to 1 (and hence $\eta_{\rm rot}$ decreases from 1 to $C_P^{\rm Rat}/C_{P,{\rm ADT}}^{\rm Rat}$) as $C_T$ increases from 0 to $C_T^{\rm Rat}$, noting that the value of $C_T/C_{P,{\rm ADT}}$ approaches 1 as $C_T$ approaches 0. The exponent `1/2' in the formulation of $\sigma$ has been determined empirically. Figure \ref{fig3} shows comparisons of this simple analytical model with `standard' $C_P$ versus $C_T$ curves for the DTU 10 MW, IEA 10 MW and IEA 15 MW reference turbines, confirming that the model agrees well with these `real' rotor performance curves for the whole range of $C_T$ up to the rated $C_T$. Note that the sudden drop of $C_P$ above the rated $C_T$ cannot be predicted by this simple analytical model, but this drop corresponds to the degradation of rotor efficiency at low wind speeds, at which the operating conditions of these rotors are not aerodynamically optimal. A very similar type of analytical rotor efficiency model has also been proposed by \citet{vdL22} and validated for a wider range of common wind turbine rotors, showing the same trend as the model proposed here.

%%%%%%%%%%%%%%%%%%%%%%%%%%%%%%%%%%%%%%%%%%%%

\section{Examples of theoretical solutions} \label{sec:results}

\subsection{Relationship between $\beta$ and $C_{PG}$}
\begin{figure}[t]
\centering{\includegraphics[width=30pc]{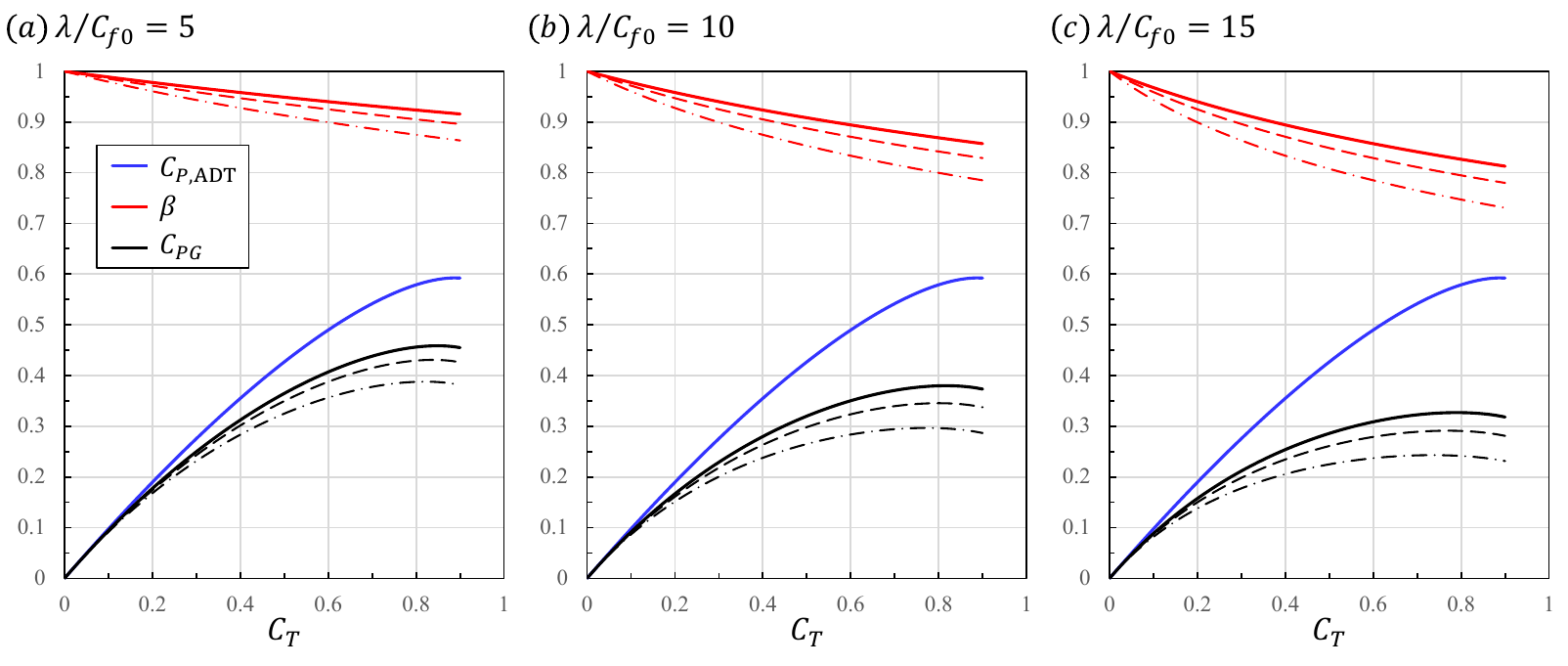}}\vspace{-2pt}
\caption{The efficiency of an idealised wind farm ($\eta_{\rm rot}=\chi_T=\chi_P=1$) with different effective array densities ($\lambda/C_{f0}$) and effective ABL heights ($h_0/LC_{f0}$). Solid lines: $h_0/LC_{f0}=20$. Dashed lines: $h_0/LC_{f0}=15$. Dash-dot lines: $h_0/LC_{f0}=10$.}
\label{fig4}
%\end{figure}
%\begin{figure}[t]
\vspace{13pt}
\centering{\includegraphics[width=30pc]{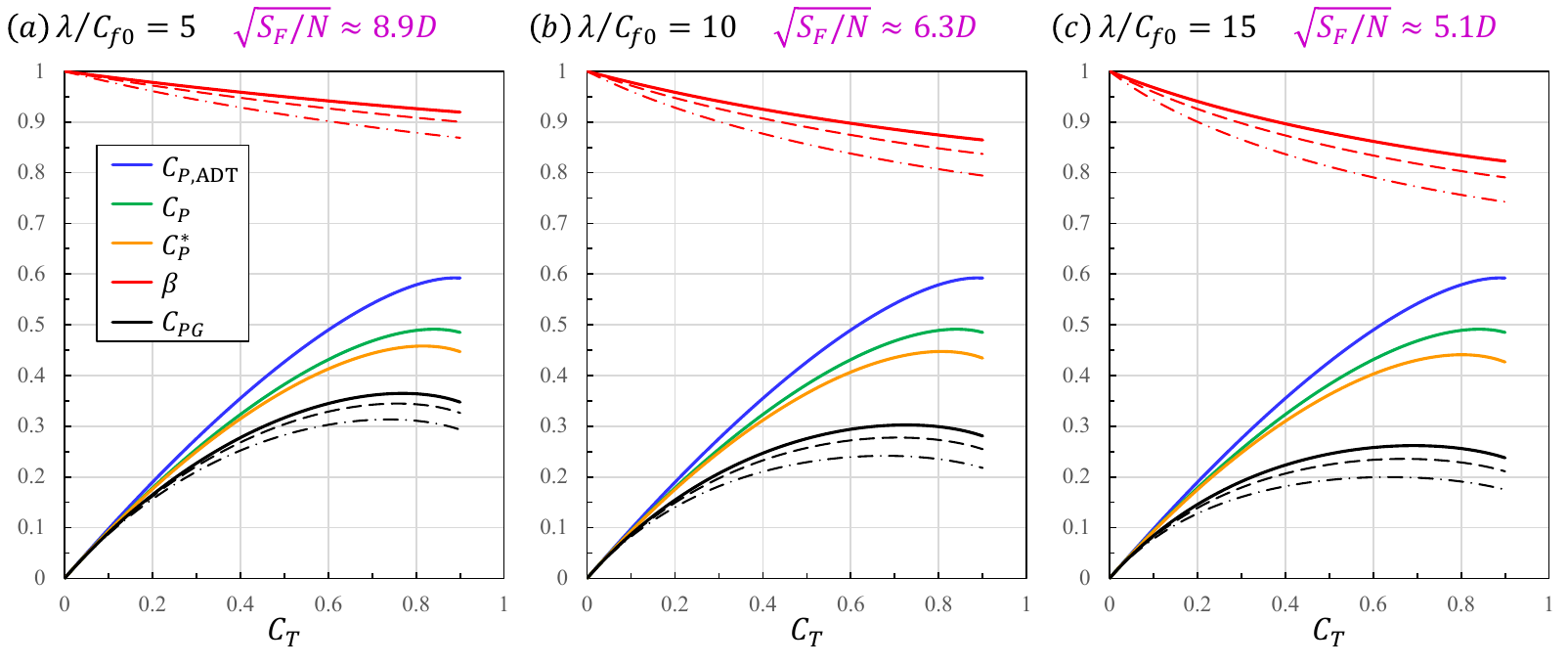}}\vspace{-2pt}
\caption{The efficiency of a realistic wind farm with IEA 15 MW turbines ($C_T^{\rm{Rat}}=0.8$, $C_P^{\rm{Rat}}=0.489$) at a typical offshore site ($C_{f0}=0.002$)  with different values of $\lambda/C_{f0}$ and $h_0/LC_{f0}$. Solid lines: $h_0/LC_{f0}=20$. Dashed lines: $h_0/LC_{f0}=15$. Dash-dot lines: $h_0/LC_{f0}=10$.}
\label{fig5}
\end{figure}

First, we present the farm wind-speed reduction factor $\beta$ (obtained from the NDFM equation (\ref{eqn:NDFM}) with the model of $M$ in (\ref{eqn:Mmodel})) and the global power coefficient $C_{PG}$ $(=\eta_{\rm ext}\eta_{\rm int}\eta_{\rm rot}C_{P,{\rm ADT}})$ in figures \ref{fig4} and \ref{fig5} for idealised and realistic farms, respectively, for a range of effective array density ($\lambda/C_{f0}=5, 10, 15$) and effective ABL height ($h_0/LC_{f0}=10, 15, 20$). Also presented together in figure \ref{fig5} are the internal power coefficient $C_P^*$ $(=\eta_{\rm int}\eta_{\rm rot}C_{P,{\rm ADT}})$ and the effective turbine power coefficient $C_P$ $(=\eta_{\rm rot}C_{P,{\rm ADT}})$. For a typical offshore case with $C_{f0}=0.002$, these $\lambda/C_{f0}$ values correspond to the average turbine spacing of about $8.9D$, $6.3D$ and $5.1D$, respectively. In addition, if we assume that the farm length $L$ is 20 km, these $h_0/LC_{f0}$ values would correspond to typical offshore ABL heights of 0.4, 0.6 and 0.8 km, respectively. Both figures demonstrate some general trends: (i) $\beta$ decreases (i.e. farm-average wind speed decreases) as $C_T$ increases, (ii) for a given wind farm, there is an optimal $C_T$ and thus an optimal $\beta$ that maximise $C_{PG}$, and (iii) for a given $C_T$, both $\beta$ and $C_{PG}$ decrease as the effective farm density ($\lambda/C_{f0}$) increases and/or the effective ABL height ($h_0/LC_{f0}$) decreases.

Note that the ratio of $C_{PG}$ to $C_{P,{\rm ADT}}$ in figure \ref{fig4} is the 'external' efficiency $\eta_{\rm ext}$ ($=\beta^3$) since there is no other power loss in these idealised cases ($C_{P,{\rm ADT}}=C_P=C_P^*$ and thus $\eta_{\rm rot}=\eta_{\rm int}=1$). For these idealised wind farms, $\eta_{\rm ext}$ is equivalent to $\eta_{FS}$ proposed by \citet{KNL25}. However, in realistic wind farms (figure \ref{fig5}) there are power losses due to rotor design (i.e. $C_{P,{\rm ADT}}>C_P$) and direct wake interference (i.e. $C_P>C_P^*$). Since $C_T^*$ also decreases from $C_T$ due to direct wake interference, $\beta$ is slightly higher in figure \ref{fig5} than in figure \ref{fig4}, meaning that $\eta_{\rm ext}$ is higher in the realistic case than in the idealised case, even though $C_{PG}$ is lower in the realistic case. It is therefore important to note that, for realistic wind farms, $\eta_{\rm ext}$ is different from $\eta_{FS}$ proposed by \citet{KNL25}. Nonetheless, the new theoretical results in figure \ref{fig5} indicate that the power loss due to `farm-atmosphere interactions' (or the reduction of the farm-average wind speed) is still likely to be the most dominant part of power losses for typical offshore wind farms. In figure~\ref{fig5} the reduction from $C_P^*$ to $C_{PG}$ (due to `external' flow interactions) is more significant than the reduction from $C_P$ to $C_P^*$ (due to `internal' flow interactions). This is similar to the conclusion drawn by \citet{KND22}.

\subsection{Iterative solution for optimal $\beta$ and $C_T$} \label{sec:semi-analytcal}

Of interest here is how the optimal $C_T$ that maximises $C_{PG}$ changes with various factors. In contrast to $C_{P,{\rm {ADT}}}$, which always takes the maximum value of $16/27\approx 0.593$ (the Betz limit) at a high optimal $C_T$ value of $8/9\approx 0.889$, the farm power is maximised at a lower optimal $C_T$ value that depends on $\lambda/C_{f0}$ and $h_0/LC_{f0}$ as well as on the rotor design and turbine layout factors. If $C_T$ is set higher than this optimal value, $C_{PG}$ becomes lower than the maximum achievable value whilst $\beta$ also decreases, meaning that the farm produces less power and creates a stronger farm-scale wake compared with the optimal case. Hence, it is of practical importance for wind farm operators to be able to predict this optimal point to ensure that $C_T$ does not exceed its optimum, or equivalently, $\beta$ does not fall below its optimum. In this subsection we derive a simple iterative method from the present theory to obtain the optimal $\beta$ and $C_T$ values, $\beta_{\rm (opt)}$ and $C_{T\rm(opt)}$, for a given wind farm.

To identify $\beta_{\rm (opt)}$ we set $\partial C_{PG}/\partial\beta=0$, which is equivalent to setting $\partial C_{PG}/\partial C_T=0$ as $\partial C_T/\partial\beta$ is generally non-zero in the vicinity of $\beta_{\rm (opt)}$. Recalling $C_{PG}=\beta^3 C_P^*$ we obtain
\begin{equation}
    \frac{\partial C_{PG}}{\partial \beta} = \frac{\partial C_{PG}}{\partial C_T}\frac{\partial C_T}{\partial \beta} = 0 \hspace{6pt} \rightarrow \hspace{6pt}  \frac{\partial C_{PG}}{\partial C_T}=3\beta^2\frac{\partial\beta}{\partial C_T}C_P^* + \beta^3\frac{\partial C_P^*}{\partial C_T} = \frac{3}{\beta}\frac{\partial\beta}{\partial C_T} + \frac{\partial C_P^*}{\partial C_T}\frac{1}{C_P^*} = 0.
    \label{eqn:CTderivatives}
\end{equation}
To solve for $C_{T\rm(opt)}$ (and by extension $\beta_{\rm (opt)}$) we must find expressions for $\partial \beta/\partial C_T$ and $\partial C_P^*/\partial C_T$ in equation~(\ref{eqn:CTderivatives}). These can be found from the NDFM equation~(\ref{eqn:NDFM}) and the new rotor efficiency and layout factor models outlined in Section~\ref{sec:newTheory}, as shown below.

By combining equations~(\ref{eqn:NDFM}) and (\ref{eqn:Mmodel}), assuming $\gamma=2$ (see Section \ref{sec:TSMT}) and noting that $C_T^*=\chi_TC_T$, we find the relationship between $\beta$ and $C_T$ as 
\begin{equation}
    \chi_T C_T\frac{\lambda}{C_{f0}} + 1  = \frac{1+\frac{h_0}{LC_{f0}}(1-\beta^2)}{\beta^3}.
    \label{eqn:momentumFull}
\end{equation}
Taking derivatives with respect to $C_T$ on both sides of this equation, we obtain
\begin{equation}
    \frac{\lambda}{C_{f0}} (\chi_T + \frac{\partial\chi_T}{\partial C_T}C_T) = \frac{\partial\beta}{\partial C_T}\left( -\frac{3(1 + h_0/LC_{f0})}{\beta^4} + \frac{h_0/LC_{f0}}{\beta^2} \right) ,
\end{equation}
which can be rearranged to obtain an expression for $\partial\beta/\partial C_T$ as
\begin{equation}
    \frac{\partial\beta}{\partial C_T} = \frac{\chi_T \lambda}{C_{f0}} \left(\frac{\beta^4}{\beta^2h_0/LC_{f0}-3(1+h_0/LC_{f0})}\right) +\Phi_{\rm non-id}C_T = \left[\frac{\partial\beta}{\partial C_T}\right]_{\rm id} +\Phi_{\rm non-id}C_T , 
    \label{eqn:derivative1}
\end{equation}
where $[\partial\beta/\partial C_T]_{\rm id}$ denotes $\partial\beta/\partial C_T$ for idealised cases without layout effects, while $\Phi_{\rm non-id}$ represents additional effects due to non-ideal turbine layout for realistic cases:
\begin{equation}
    \Phi_{\rm non-id} = \frac{\partial\chi_T}{\partial C_T}\frac{\lambda}{C_{f0}}\left(\frac{\beta^4}{\beta^2h_0/LC_{f0}-3(1+h_0/LC_{f0})} \right),
    \label{eqn:non-ideal4}
\end{equation}
which is zero for idealised cases since $\partial\chi_T/\partial C_T=0$. For realistic cases, $\partial\chi_T/\partial C_T$ can be evaluated using the layout factor model in equation~(\ref{eqn:chimodel}), which gives $\partial\chi_T/\partial C_T=2\chi \partial\chi/\partial C_T$ and
\begin{equation}
    \frac{\partial\chi}{\partial C_T} = -\frac{1}{2}C_\chi\frac{1}{\sqrt{1-C_T}}\frac{1}{(1+2k\sqrt{\pi/4\lambda})^2}. 
    \label{eqn:chiDerivative}
\end{equation}
Hence, we have an expression for the first term in equation~(\ref{eqn:CTderivatives}), given initial guesses for $\beta$ and $C_T$. For the second term in equation~(\ref{eqn:CTderivatives}), recalling $C_P^*=\chi_P\eta_{\rm rot}C_{P,{\rm ADT}}$ we obtain
\begin{equation}
    \frac{\partial C_P^*}{\partial C_T}\frac{1}{C_P^*} = \frac{\partial [{\rm ln}(C_P^*)]}{\partial C_T} = \frac{1}{C_T} - \frac{1}{2}\frac{1}{(1-C_T+\sqrt{1-C_T})} + \Psi_{\rm non-id} ,
    %\frac{\partial }{\partial C_T}\left[ln(\chi_P) + ln(\eta_{rot}) + ln(1/2) + ln(C_T) + ln(1 + \sqrt{1-C_T})\right]
    \label{eqn:derivative2}
\end{equation}
where $\Psi_{\rm non-id}$ represents the effects of non-ideal turbine layout and rotor design:
\begin{equation}
    \Psi_{\rm non-id} = \frac{\partial\chi_P}{\partial C_T}\frac{1}{\chi_P} + \frac{\partial\eta_{\rm rot}}{\partial C_T}\frac{1}{\eta_{\rm rot}} ,
    \label{eqn:psi}
\end{equation}
which is zero for idealised farms, similarly to $\Phi_{\rm non-id}$. For realistic farms, the first term of $\Psi_{\rm non-id}$ can be evaluated using the layout factor model in equation~(\ref{eqn:chimodel}), which gives $\partial\chi_P/\partial C_T=3\chi^2 \partial\chi/\partial C_T$ and equation~(\ref{eqn:chiDerivative}). The second term of $\Psi_{\rm non-id}$ can be evaluated using equation~(\ref{eqn:CPmodel}), which gives
\begin{equation}
    \frac{\partial\eta_{\rm rot}}{\partial C_T} = \frac{\partial \sigma}{\partial C_T}\left( \frac{C_P^{\rm Rat}}{C_{P,{\rm ADT}}^{\rm Rat}} -1 \right), \hspace{10pt} \frac{\partial \sigma}{\partial C_T} = \frac{1}{2\sigma}\left(\frac{1}{\frac{C_T^{\rm Rat}}{C_{P,{\rm ADT}}^{\rm Rat}} - 1}\right)\frac{1}{\sqrt{1-C_T}(1+\sqrt{1-C_T})^2} .
    \label{eqn:non-ideal2}
\end{equation}
Hence, we have an expression for $\Psi_{\rm non-id}$ in equation~(\ref{eqn:derivative2}) and thus an expression for the second term in equation~(\ref{eqn:CTderivatives}), given initial guesses for $\beta$ and $C_T$.

Finally, to iteratively solve equation~(\ref{eqn:CTderivatives}), we insert equations~(\ref{eqn:derivative1}) and~(\ref{eqn:derivative2}) and rearrange to form a cubic equation in $C_T$:
\begin{equation}
    -\gamma_1C_T^3 + [\gamma_1(1+\gamma_3)-\gamma_2]C_T^2 + \left[\gamma_2(1+\gamma_3)-\frac{3}{2}\right]C_T + (1+\gamma_3)=0,
    \label{eqn:cubicCT2}
\end{equation}
where the coefficients are given by
\begin{equation}
     \gamma_1=\frac{3}{\beta} \Phi_{\rm non-id}, \hspace{12pt}  \gamma_2=\frac{3}{\beta}\left[\frac{\partial\beta}{\partial C_T}\right]_{\rm id} + \Psi_{\rm non-id}, \hspace{12pt} \gamma_3 = \sqrt{1-C_T}.
    \label{eqn:coeffs2}
\end{equation}
The cubic equation~(\ref{eqn:cubicCT2}) can be solved with its coefficients $\gamma_1$, $\gamma_2$ and $\gamma_3$ being evaluated from initial guesses of $C_T$ and $\beta$, which should be within the realistic range of 0 to 1. The resulting value of $C_T$ obtained by solving the cubic equation (\ref{eqn:cubicCT2}) is then used to find $\beta$ by solving the cubic equation~(\ref{eqn:momentumFull}). These new values for $C_T$ and $\beta$ are then used to find $\gamma_1$, $\gamma_2$ and $\gamma_3$, and the cubic equation~(\ref{eqn:cubicCT2}) is solved again with the updated coefficients, and the process is repeated until $C_T$ and $\beta$ converge. The resulting values are $C_{T\rm(opt)}$ and $\beta_{\rm (opt)}$ required to achieve the maximum global power coefficient, $C_{PG(\rm max)}$. 

Note that in the present study the `non-ideal' terms $\Phi_{\rm non-id}$ and $\Psi_{\rm non-id}$ were evaluated using the simple analytical models for the layout effect and the rotor efficiency introduced in Section~\ref{sec:newTheory}. However, these terms could also be evaluated directly from more accurate data (e.g. wind farm LES and turbine performance data) and the iterative method presented above would still apply, requiring only revised expressions for the non-ideal terms $\Phi_{\rm non-id}$ and $\Psi_{\rm non-id}$.

\begin{figure}[t]
\centering{\includegraphics[width=34pc]{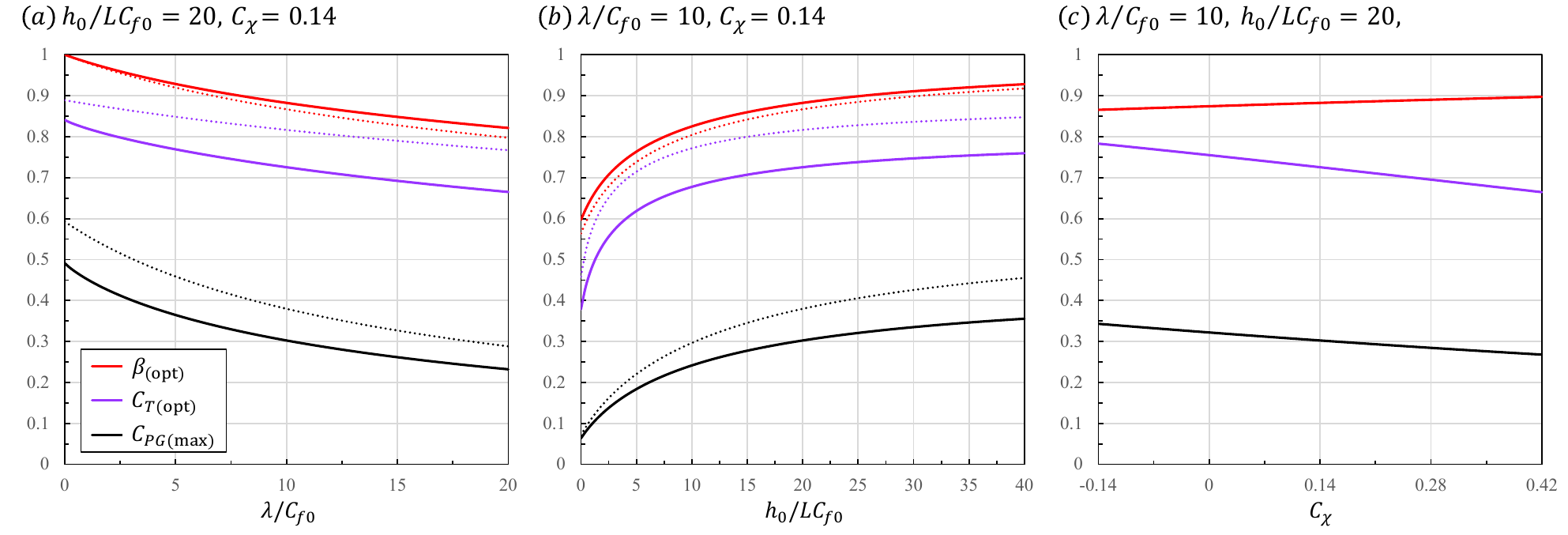}}\vspace{-0pt}
\caption{The maximum global power coefficient $C_{PG({\rm max})}$, optimal wind speed reduction factor $\beta_{({\rm opt})}$ and optimal turbine thrust coefficient $C_{T({\rm opt})}$ for typical ranges of $\lambda/C_{f0}$, $h_0/LC_{f0}$ and $C_{\chi}$. Solid lines: realistic offshore wind farm ($C_{f0}=0.002$) with IEA 15 MW turbines ($C_T^{\rm{Rat}}=0.8, C_P^{\rm{Rat}}=0.489$). Dotted lines: idealised wind farm ($\eta_{\rm rot}=\chi_T=\chi_P=1$).}
\label{fig6}
\end{figure}

Figures~\ref{fig6}($a$) and ($b$) show how $\beta_{({\rm opt})}$, $C_{T({\rm opt})}$ and $C_{PG({\rm max})}$ vary with $\lambda/C_{f0}$ (at a typical effecitive ABL height of $h_0/LC_{f0}=20$) and $h_0/LC_{f0}$ (at a typical effective array density of $\lambda/C_{f0}=10$), for idealised and realistic wind farms. As expected from the performance curves in figures \ref{fig4} and \ref{fig5}, both $C_{T({\rm opt})}$ and $\beta_{({\rm opt})}$ decrease as $\lambda/C_{f0}$ increases and/or $h_0/LC_{f0}$ decreases. Importantly, $C_{T({\rm opt})}$ is lower for the realistic farm than for the idealised farm, whereas $\beta_{({\rm opt})}$ is slightly higher for the realistic farm than for the idealised farm. Note that $C_{T({\rm opt})}$ for the realistic case in figure \ref{fig6}($a$) goes above $C_T^{\rm Rat}=0.8$ at $\lambda/C_{f0}<2.3$, which appears incorrect, but this is because the analytical rotor model gives a slightly higher $C_P$ value than $C_P^{\rm Rat}=0.489$ at $C_T>C_T^{\rm Rat}$ as shown in figure \ref{fig3}($c$). It should also be noted that these realistic farm results depend on the parameter $C_{\chi}$ set in the layout factor model. Figure~\ref{fig6}($c$) shows the sensitivity of the results to $C_{\chi}$ (for a representative case with $\lambda/C_{f0}=10$ and $h_0/LC_{f0}=20$). As $C_{\chi}$ increases (implying that the wind direction is more aligned with the axes of the turbine array) both $C_{PG({\rm max})}$ and $C_{T({\rm opt})}$ decrease while $\beta_{({\rm opt})}$ increases, resulting in a higher $\eta_{\rm ext}$ and a lower $\eta_{\rm int}$.

\subsection{Maximum farm power and power density}
The negative relationship between $C_{PG{\rm (max)}}$ and $\lambda/C_{f0}$ shown in figure \ref{fig6}($a$) means that the maximum achievable power `per turbine' decreases as we increase the number of turbines in a given farm area. However, the power density (i.e. power per unit farm area) still increases with the number of turbines, suggesting that the slope of $C_{PG{\rm (max)}}$ versus $\lambda/C_{f0}$ becomes a key factor (together with other factors such as the cost of each turbine) when deciding on the optimal number of turbines. 

\begin{figure}[t]
\centering{\includegraphics[width=28pc]{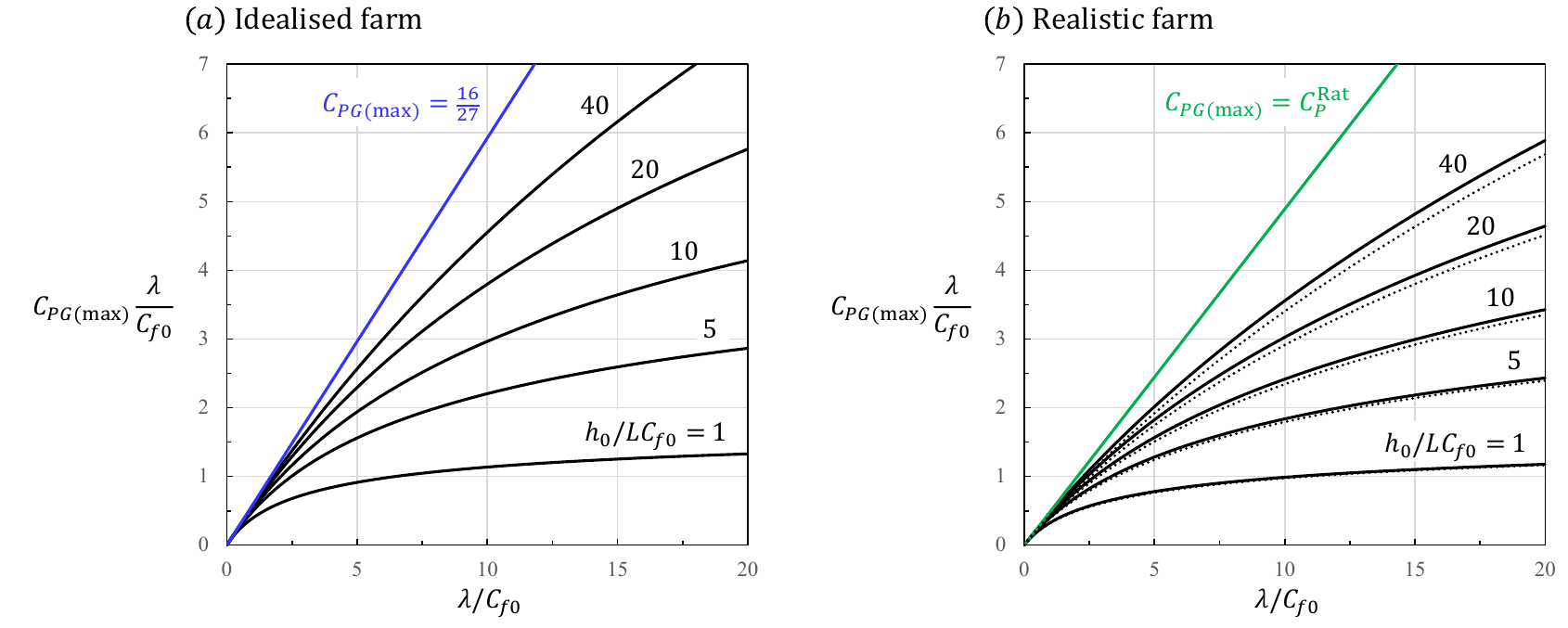}}\vspace{-2pt}
\caption{The dependence of the non-dimensional maximum power density $C_{PG({\rm max})}\lambda/C_{f0}$ on the effective array density at different effective ABL heights: (a) idealised wind farm ($\eta_{\rm rot}=\chi_T=\chi_P=1$), and (b) realistic wind farm with IEA 15 MW turbines ($C_T^{\rm{Rat}}=0.8, C_P^{\rm{Rat}}=0.489$) at a typical offshore site (solid lines, $C_{f0}=0.002$) and onshore site (dotted lines, $C_{f0}=0.01$).}
\label{fig7}
%\end{figure}
%\begin{figure}[t]
\vspace{14pt}
\centering{\includegraphics[width=34pc]{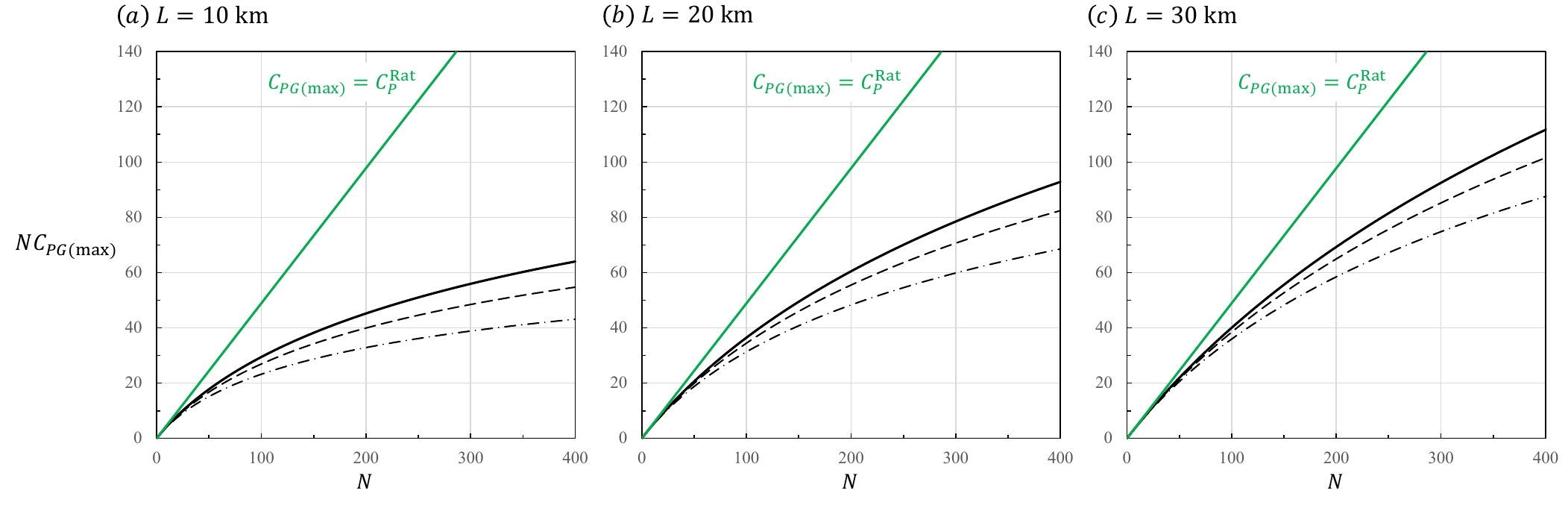}}\vspace{-3pt}
\caption{The dependence of the non-dimensional maximum farm power $NC_{PG({\rm max})}$ on the number of turbines (IEA 15 MW, $C_T^{\rm{Rat}}=0.8, C_P^{\rm{Rat}}=0.489$) for an offshore wind farm ($C_{f0}=0.002$) with three different farm sizes: (a) $L=10$ km, (b) $L=20$ km, and (c) $L=30$ km (where $S_F/A=L^2/0.04$ is assumed). Solid lines: $h_0=0.8$ km. Dashed lines: $h_0=0.6$ km. Dash-dot lines: $h_0=0.4$ km.}
\label{fig8}
\end{figure}

Figure \ref{fig7} shows how the non-dimensional maximum power density $C_{PG({\rm max})}\lambda/C_{f0}$ changes with $\lambda/C_{f0}$ and $h_0/LC_{f0}$ for idealised and realistic wind farms. For the idealised case, the non-dimensional maximum power density is determined solely by $\lambda/C_{f0}$ and $h_0/LC_{f0}$ (figure \ref{fig7}$a$). For the realistic case, this is still largely determined by $\lambda/C_{f0}$ and $h_0/LC_{f0}$, although there is also a weak dependence on $C_{f0}$ itself because the turbine layout factors $\chi_T$ and $\chi_P$ depend on $\lambda$ instead of $\lambda/C_{f0}$, resulting in small differences between the offshore and onshore cases in figure \ref{fig7}($b$). However, these non-dimensional performance curves for offshore and onshore farms must be interpreted with care, since $C_{f0}$ affects both $\lambda/C_{f0}$ and $h_0/LC_{f0}$, e.g. having a five-times higher $C_{f0}$ value at onshore means that both $\lambda/C_{f0}$ and $h_0/LC_{f0}$ are five times smaller for a given array density $\lambda(\equiv NA/S_F)$ and for a given ABL height $h_0$. Moreover, the farm length $L$ also affects both $\lambda/C_{f0}$ and $h_0/LC_{f0}$ since the farm area $S_F$ should be roughly proportional to $L^2$ (depending on the shape of the farm area; see figure \ref{fig1}).

Figure \ref{fig8} presents the power characteristics for the realistic offshore case in figure \ref{fig7}($b$) in a different (more practical) manner, to show how the non-dimensional maximum farm power $NC_{PG({\rm max})}$ changes with the number of turbines $N$ for three different sizes of wind farms (with $L=10$, 20 and 30 km) and three different ABL heights ($h_0=0.4$, 0.6 and 0.8 km). Note that the green line in each graph shows the `no flow interaction' scenario ($\eta_{\rm int}=\eta_{\rm ext}=1$) where the farm power is $\frac{1}{2}\rho U_{F0}^3 ANC_{P}^{\rm Rat}$, which is $N$ times the rated turbine power (15 MW in this case) when $U_{F0}$ is at the rated wind speed. This means that, when $U_{F0}$ is at the rated wind speed, this green line represents the scenario where the capacity factor (CF) is fixed at 100$\%$. Hence, these theoretical results highlight how quickly the CF decreases as we increase the number of turbines, especially in a small wind farm with a low ABL height (e.g. when $h_0=0.4$ km and $U_{F0}$ is at the rated wind speed, the maximum CF of this offshore wind farm with $L=10$ km and $N=200$ is only about $34\%$, as can be seen in figure \ref{fig8}a). Although not shown here for brevity, these power curves can also be predicted very quickly for different rotor design scenarios, opening the possibility of optimising rotor design and farm design in a coupled manner.

%%%%%%%%%%%%%%%%%%%%%%%%%%%%%%%%%%%%%%%%%%%%

\section{Conclusions}
In this study we have extended the two-scale momentum theory for wind farm aerodynamics \citep{ND20,KDN23} to account for power losses due to non-ideal rotor design and turbine layout effects. Most importantly, we have introduced the turbine layout factors $\chi$, $\chi_T$ and $\chi_P$ in equation (\ref{eqn:chi}), allowing us to properly differentiate the `internal' power loss (due to direct turbine-wake interference, reducing the inflow speed for individual turbines relative to the farm-average wind speed) from the `external' power loss (due to farm-atmosphere interaction, reducing the farm-average wind speed). We then proposed simple analytical sub-models for the rotor efficiency and the layout factors to provide a holistic view of the power performance of realistic wind farms. We also presented an iterative method based on the extended theory to quickly calculate the optimal farm induction factor that maximises the farm power for a given set of conditions, including the ABL height. Finally, we used the theory to demonstrate how the maximum power of a realistic offshore wind farm would change with the farm size and the number of turbines at different ABL heights.

The farm power characteristics defined and predicted by the two-scale momentum theory, such as the relationship between the farm power and the farm induction factor, are as fundamentally important for wind power generation as the turbine power characteristics defined and predicted by the classical BEM theory. Just like the BEM theory forms the basis for modern wind turbine design, the two-scale momentum theory has the potential to form the basis for future wind farm design. To further validate the two-scale theory not only for idealised wind farms (as was done by \citet{KNL25}) but also for realistic wind farms, we would need LES of large wind farms with realistic turbine models (actuator-line models, for example) in future studies. It would also be important to further improve the accuracy and applicability of the analytical sub-models in the theory, especially those for the momentum availability factor $M$ and the turbine layout factors $\chi_T$ and $\chi_P$, in future studies.

%-----\clearpage

%%%%%%%%%%%%%%%%%%%%%%%%%%%%%%%%%%%%%%%%%%%%

\begin{Backmatter}

\paragraph{Acknowledgements}
The authors thank Dr Andrew Kirby for providing his LES data from \citet{KND22}, which were used by TN to create figure \ref{fig2}.

\paragraph{Funding Statement}
This research has been partially supported by the UK Met Office (WCSSP programme, SA24-2.4) and the UK Natural Environment Research Council (NERC ECOFlow programme, NE/Z504099/1).

\paragraph{Declaration of Interests}
The authors declare no conflict of interest.

\paragraph{Author Contributions}
TN proposed the power loss mechanisms in Section \ref{sec:powerloss} and the rotor efficiency model in Section \ref{sec:phiP}. TN and ASMS jointly developed the layout factor model in Section \ref{sec:chimodel}. ASMS derived the iterative method in Section \ref{sec:semi-analytcal}.

\paragraph{Data Availability Statement}
A set of Matlab codes for the iterative method described in Section \ref{sec:semi-analytcal} will be made available in the GitHub repository (after the acceptance of this paper for publication).
 
%\paragraph{Ethical Standards}
%The research meets all ethical guidelines, including adherence to the legal requirements of the study country.

%\paragraph{Supplementary Material}
%Methods section and Supplementary information are available at \url{https://doi.org/10.1017/flo.2021.1}.

%%%%%%%%%%%%%%%%%%%%%%%%%%%%%%%%%%%%%%%%%%%%

%\appendix

%%%%%%%%%%%%%%%%%%%%%%%%%%%%%%%%%%%%%%%%%%%%

\end{Backmatter}

\end{document}